# A graphene-based electrochemical switch

T.J. Echtermeyer[1,a], M.C. Lemme[1,b], M. Baus[1], B. N. Szafranek[1],

A.K. Geim[2], H. Kurz[1]

[1]Advanced Microelectronics Center Aachen (AMICA), AMO GmbH, Otto-Blumenthal-Str. 25, 52074 Aachen, Germany

[2]Manchester Centre for Mesoscience and Nanotechnology, University of Manchester, Manchester, M13 9PL, UK

[a] echtermeyer@amo.de , +49 241 8867 229

[b] lemme@amo.de , +49 241 8867 207

**Conventional field effect transistor operation in graphene is limited by its zero gap and minimum quantum conductivity. In this work, we report on controlled electrochemical modification of graphene such that its conductance changes by more than six orders of magnitude, which enables reversible bipolar switching devices. The effect is explained by a chemical reaction of graphene with hydrogen ($H^+$) and hydroxyl ($OH^-$), which are catalytically generated from water molecules in the sub-stochiometric silicon oxide gate dielectric. The reactive species attach to graphene making it nonconductive but the process can subsequently be reversed by short current pulses that cause rapid local annealing. We believe that the demonstrated electrochemical field effect devices are viable candidates for future logic circuits, non-volatile memories and novel neuromorphic processing concepts.**

The current exponential growth of interest in graphene and its applications can – to a large extent – be attributed to its remarkable electronic properties that include carrier mobilities in excess of 20,000 cm$^2$/Vs and a submicron mean free path at room temperature (*1,2*). One of the most discussed research directions on graphene is probably its use in nanoelectronics as the base material for field effect transistor (FET) type applications (*2-8*). However, a major hurdle remaining here is graphene's minimum conductivity (even at the neutrality (Dirac) point where no carriers are nominally present) (*2*), which results in low $I_{on}/I_{off}$ ratios insufficient for most electronic applications. Indeed, the best ratio achieved to date with a top gated graphene FET at room temperature is only six (*8*). It is generally believed that this problem could be circumvented by using graphene nanoribbons (GNRs) narrower than 5 nm, which in principle can open up a band gap larger than 500 meV (*9,10*). However, even state-of-the-art nanolithography tools do not allow reliable control at this spatial scale (*11*), and the smallest GNRs demonstrated so far reached widths of ~10 nm (*2,7,12*), with electrical data available only for ~30nm GNRs showing a semiconductor FET behavior at liquid-helium temperatures. Therefore, it remains a formidable challenge to demonstrate GNR-type transistors that can be used in integrated circuits, especially without loosing graphene's electronic quality.

In this report, we demonstrate graphene field effect devices (FEDs) that rely on an electrochemical mechanism and allow $I_{on}/I_{off}$ ratios $>10^6$ at room temperature. We attribute their transistor action to attachment/detachment of active radicals to/from graphene in reduction-oxidation-like (redox) processes that strongly alter graphene's electronic properties.

Our graphene FEDs (we use this notion to indicate a different operational mechanism with respect to the conventional FETs) were fabricated by micromechanical cleavage of graphite (*1*) on top of a silicon substrate with 300 nm of silicon dioxide ($SiO_2$). We also used a 20-nm layer of silicon oxide ($SiO_x$) evaporated on top of graphene as a top-gate dielectric and a 40 nm tungsten film for source, drain and top-gate electrodes.

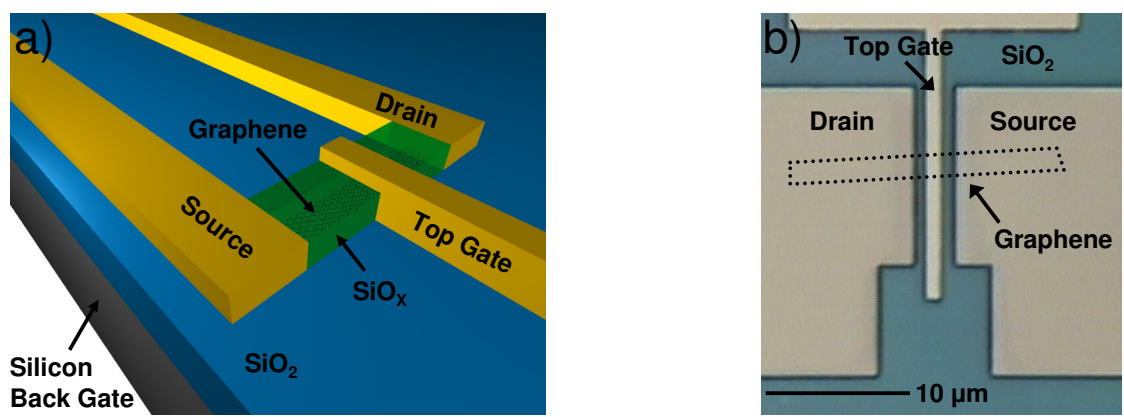

**Fig. 1: Graphene-based electrochemical devices. a) Schematic view of the double-gated devices used in the experiments. A global back gate and a local top gate electrodes controlled graphene's conductance. b) Optical micrograph of one of our devices.**

Figure 2 shows typical top gate transfer characteristics (drain current $I_d$ versus top gate voltage $V_{tg}$) measured in our devices under ambient conditions. First, the gate voltage was swept back and forth between $V_{tg}$ = -4 V and $V_{tg}$ = 4 V (open symbols). One can see that on a logarithmic scale the measured curves were practically flat, exhibiting a current level in the µA range (channel resistance of ≈10kΩ). We refer to this metallic state of FEDs as an on-state. Note that when plotted on a linear scale (see supporting online material), the transfer characteristics showed the standard "V-shaped" behavior expected for two-terminal graphene devices with low $I_{on}/I_{off}$ ratios.
In the next measurement in Fig. 2, we extended the range of top-gate voltage, above a certain breakdown value that was found to be ≈±5V for 20nm $SiO_2$. After such a voltage was applied, the drain current $I_d$ fell down by over seven orders of magnitude, indicating an insulating state (accordingly, called an "off-state"). The channel resistance in the off-state exceeded 10GΩ, which was comparable to a typical leakage current in our experimental setup so that devices with discontinuous graphene layers also exhibited the same resistance level. However, our FEDs were not destroyed in the off-state as seen in Fig. 2 (solid symbols). Indeed, if we swept voltage starting from the insulating state at, say, $V_{tg}$ = -5 V towards $V_{tg}$ = 5 V, the device remained insulating for negative voltages but, around $V_{tg}$ = 0 V, the drain current suddenly recovered to ≈60% of its initial on-state value. Finally, as the gate voltage exceeded +4 V, FEDs switched again into the insulating state. When the sweep direction was reversed so that we started at $V_{tg}$ = +5 V (Fig. 2, squares), the switching effect was essentially mirrored: the device remained in the off-state for positive voltages and recovered to the metallic state at negative $V_{tg}$. For top-gate voltages beyond -3V, the device turned off again. Consecutive measurement cycles (not shown in Fig. 2) revealed similar behavior but the degree of recovery gradually decreased with increasing the number of cycles.

We attribute the observed gate-induced switching to an electrochemical modification of the graphene structure. Indeed, two very relevant examples of such modification are known from literature. One is so-called graphane, that is, a graphene derivative with hydrogen atoms attached in $sp^3$ configuration. A recent theory study predicted a band gap in graphane of 3.5 eV (*13*). Another example is graphene oxide that is believed to be a graphene sheet with a massive amount of hydroxyl or similar groups attached to the surface, which makes it insulating at room temperature (*14*,*15*). To this end, it was also shown that one $sp^3$ carbon–oxygen or carbon–hydroxyl bond per $10^6$ $sp^2$ bonds decreased the conductivity of CNTs by 50% (*16*). Therefore, it is reasonable to suggest that the underlying physics mechanism of the observed switching is an electrostatic generation of hydroxyl or hydrogen in the graphene FEDs and their following electrochemical reaction (chemisorption) with graphene.

This assumption is in agreement with the bipolar nature of the switching events, which implies that both negatively charged hydroxyl (OH⁻) and positively charged hydrogen (H⁺) are involved.

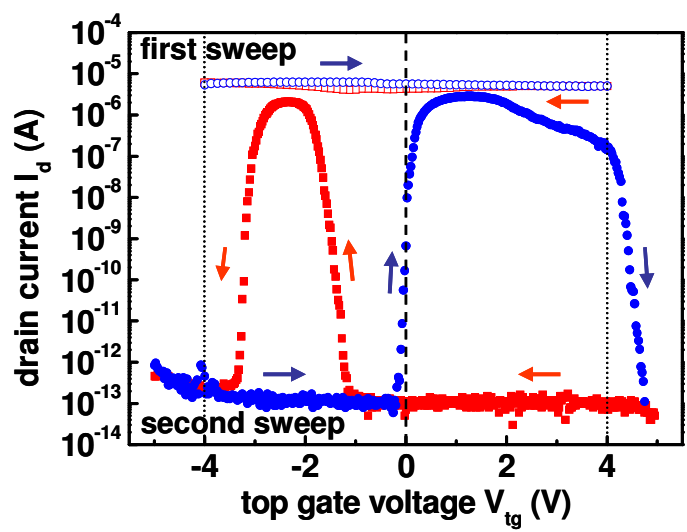

**Fig. 2: Graphene's conductance switching. The top curves (open symbols) show typical top-gate transfer characteristics for our FEDs (drain-source voltage $V_{ds}$ =50 mV). The device was also back-gated to its neutrality point by applying a back-gate voltage $V_{bg}$ =-15V. In the second sweep, we used slightly higher top-gate voltages and could switch the drain current on and off. The effect is clearly bipolar, being observed in both sweep directions.**

To prove the importance of water (as a source for hydrogen and hydroxyl), we have carried out a series of controlled-environment experiments (see supporting online material). Figure 3 shows the influence of environment on the drain current. Here we exposed a device sequentially to vacuum, nitrogen ($N_2$) and water vapor (controlled but below the ambient moisture level) for extended periods of time at different top-gate voltages. Prior to each time sequence in a different environment the back-gate transfer characteristics were measured, and the range of achieved drain currents is indicated by the shaded areas in figure 3. The minimum current corresponds to the neutrality point. In vacuum (Fig. 3a), the device showed no time dependence as its resistance remained practically constant. Only small changes in $I_d$ (≈0.05μA) with $V_{tg}$ were observed due to the conventional electric field effect (*2*). When nitrogen was let into the chamber, the drain current became slightly time dependent (Fig. 3b). In addition, the top gate voltage caused distinctly more pronounced changes as compared to the vacuum environment. We attribute them to gradual electrochemical doping of graphene, probably by remnant active molecules present in or transferred through the chamber by inert $N_2$ (*17*). Finally, when water vapor was introduced (Fig. 3c), we observed a strong decrease in $I_d$ for top gate voltages exceeding -4 V. In water vapor, the influence of $V_{tg}$ was clearly stronger as compared to $N_2$ and especially vacuum. In addition, repetitive sweeps at $V_{tg}$ = -5.5 V revealed that the effect was cumulative rather than just a function of $V_{tg}$. Furthermore and in contrast to vacuum and nitrogen environment, the current for top gate voltages ≤ -5 V was well below its lowest (neutrality-point) value found in the back-gate measurements. This observation rules out the simple chemical doping (*17*) that only shifts the $I_d$ minimum but cannot make graphene non-metallic, with conductivity below that reached at the neutrality point (*2*). We attribute the strong decrease in $I_d$ observed at high $V_{tg}$ to the onset of the insulating state induced by the electrochemical doping that involves graphene oxidation and is amplified by transverse electric field. In this experiment, the water concentration was less than under the ambient conditions (see online supporting material), which did not allow us to reach the fully developed insulating state at -5V.

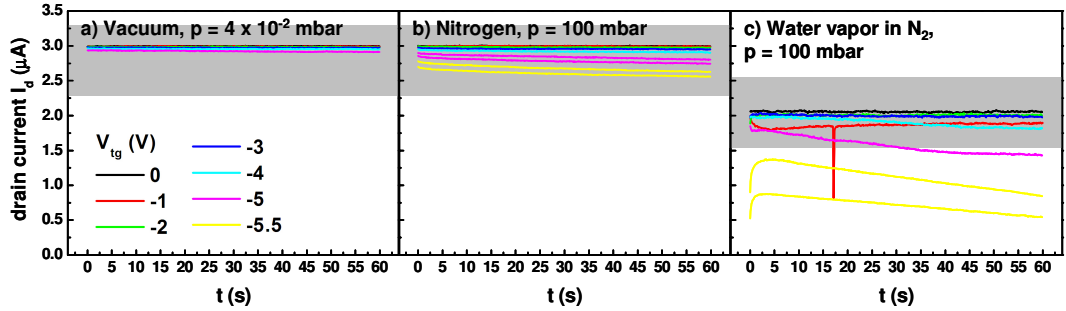

**Fig. 3: Importance of the environment. Time dependence of the drain current measured repeatedly for 1 min intervals after stepwise changes of $V_{tg}$ from 0 to -5.5 V in vacuum (a), nitrogen (b) and water vapor in $N_2$ (c). $V_{ds}$ =50 mV and $V_{bg}$ =0V. The presence of water vapor induced a major decrease in the device conductance at high $V_{tg}$. The grey-shaded areas show the range of $I_d$ achieved by applying back-gate voltages (range ± 100V), before measuring each of the $V_{tg}$ sequences. The lower boundary corresponds to the drain current at the neutrality point.**

The distinct influence of the environment on the FED characteristics was observed despite our graphene films were completely covered by a 20 nm thick silicon oxide layer. This is inline with literature, which shows that $H_2O$ and its products are highly mobile in $SiO_x$ layers (*18,19*). Moreover, silicon-rich $SiO_x$ has been known to act as a catalyst to break up $H_2O$ into $H^+$ and $OH^-$ at an energy cost of only 0.3 eV (*18*). This further supports our model for the observed electrochemical switching, which is graphically illustrated in figure 4.

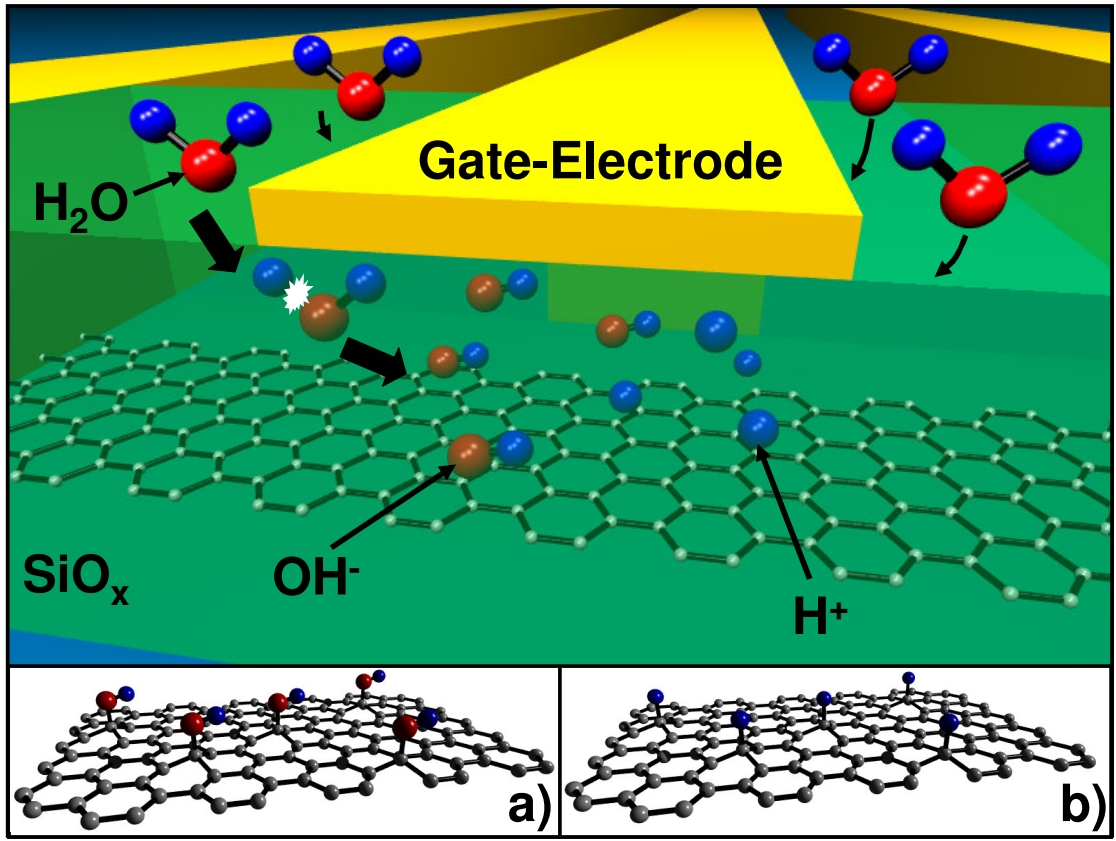

**Fig. 4: Electrochemical modification of graphene.** Water from the environment is attracted by the electrostatic pressure inside the capacitor formed by the top gate, silicon-oxide dielectric and graphene. Inside the porous oxide layer, $H_2O$ molecules are then broken up into $H^+$ and $OH^-$ by electric field, and the effect is strongly enhanced by the presence of catalytically-active $SiO_X$. Depending on field polarity, the reactive ions diffuse towards graphene and chemically (by $sp^3$ bonding) transform it into an insulating derivative such as graphane or graphene oxide. Inset: schematic of transitions towards graphene oxide (a) and graphane (b).

In order to utilize the observed switching effect in viable future applications, it is obviously essential to find reliable ways of controlling the switching and repeating the cycle numerous times. In particular, this concerns reversibility of graphene FEDs into the original metallic state without significant aging effects. To this end, we have employed short current pulses of 80 µs in length (the shortest pulses allowed by our experimental setup) to restore the metallic conductivity of graphene from its insulting state. Fig. 5a shows the output characteristics $I_d(V_{ds})$ of one of our FEDs, measured again under ambient conditions. Here, in the initial state, the drain current was in the micro-amp range. By applying a top gate voltage of $V_{tg}$ =-5 V, the drain current was quenched – similar to Fig. 2 – by more than a million times into the pA range (off-state; Fig. 5a). After reaching the insulating state, a current pulse of $I_d$ = 20 µA was supplied through the source and drain electrodes. This has lead to a partial recovery of the current level by two orders of magnitude. Subsequent pulses of 20 µA and 40 µA improved the device conductance only marginally (by 10 times) so that it eventually saturated and the metallic state could not be reached. However, when a current pulse of $I_d$ = 50 µA was supplied, our FED showed a complete recovery and the initial on-state was recovered. We suggest that the graphene film is heated by the current pulses (see supporting online material) and, when a critical temperature is achieved by supplying a sufficient current, the rapid annealing completely removes attached $sp^3$-bonding radicals from the graphene surface. This observation is in broad agreement with the fact that shallow dopants can be removed from doped (metallic) graphene samples by their thermal annealing in vacuum or inert atmosphere (see, e.g., (17)) and, also, by supplying an electric current directly through graphene devices (20). In our case, we remove much stronger bounded species responsible for the insulating state (chemisorption instead of physisorption).

The two distinct resistance states (metallic and insulating) observed in Fig. 5a may well be interpreted as ON and OFF states in a solid state switch or memory. To prove the concept of such graphene-based electronic switching devices further, Fig. 5b shows repetitive ON-OFF cycles for the same device. Among the controlled switching events shown here, the first eight correspond to the experiment in Fig. 5a. In the following measurements, the metallic ON state was reliably and repeatedly reached by a single 50 µA current pulse. The insulating OFF state (grey bars) was also reached reliably, albeit there was a margin of several orders of magnitude (with resistances between tens of MΩ and GΩ), which probably indicates a different level of oxidation and requires further investigation. Nonetheless, our initial experiments clearly demonstrate the potential of electrochemical graphene switches. In this respect, let us note that the observed resistance changes of more than six orders of magnitude (and so far limited by current leakage) are much larger than in comparable "resistive" switches, such as phase-change memories (PCRAM) (*21*). In addition, Fig. 5 suggests that the range of achieved resistance values in graphene FEDs is sufficiently large to potentially allow ternary and even multi-state logic.

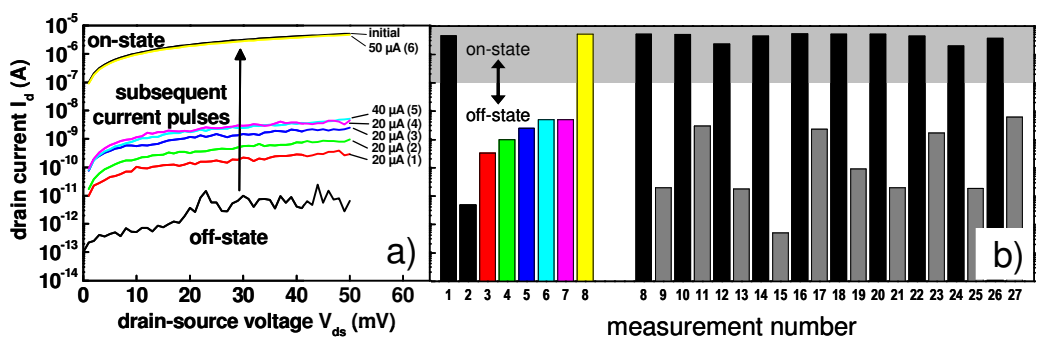

**Fig. 5: Graphene-based electronic switch. a) Output characteristics of a graphene FED in different resistance states. By applying a top-gate voltage exceeding -5V, we set the device in the off state. A short current pulse of 50 µA led to a complete recovery of the drain current. b) Repetitive ON-OFF switching of the FED by applying top-gate voltages and current pulses. The height of**

**each bar shows the drain current at $V_{ds}$ = 50 mV for different states of the device.**

The demonstrated electrochemical switching of graphene's conductivity with its distinct on- and off-states has a clear potential for future (non-volatile) memory and logic applications. Our experiments have for the first time exploited the fact that graphene is robust and sufficiently inert but, at the same time, being one atom thick, fully open to controllable (electro)chemical modification of its structure and electronic properties. Further improvements in switching times, level accuracy and reliability can reasonably be expected as better understanding of the involved processes is achieved. In a long term perspective, the presence of intermediate states may open the possibility for neuromorphic processors and networks.

22. The authors would like to thank J. Bolten and T. Wahlbrink for their e-beam lithography support. T.J. Echtermeyer thanks A. Dybek, P. von Doetinchem and J.C. Meyer for enlightening discussions. Financial support by the German Federal Ministry of Education and Research (BMBF) under contract number NKNF 03X5508 ("Alegra") is gratefully acknowledged.